\documentclass[default]{sn-jnl}

\usepackage{natbib}

\begin{document}
\title[Transcriptomics-based matching of drugs to diseases with deep learning]{Transcriptomics-based matching of drugs to diseases with deep learning}

\author*[1]{\fnm{Yannis} \sur{Papanikolaou}}\email{yannis.papanikolaou@healx.io}
\author[1]{\fnm{Francesco} \sur{Tuveri}}\email{francesco.tuveri@healx.io}
\author[1]{\fnm{Misa} \sur{Ogura}}\email{misa.ogura@healx.io}
\author[1]{\fnm{Daniel} \sur{O'Donovan}}\email{dan.odonovan@healx.io}

\affil[1]{\orgname{Healx}, \city{Cambridge}, \country{UK}}


\abstract{In this work we present a deep learning approach to conduct hypothesis-free, transcriptomics-based matching of drugs for diseases. Our proposed neural network architecture is trained on approved drug-disease indications, taking as input the relevant disease and drug differential gene expression profiles, and learns to identify novel indications. We assemble an evaluation dataset of disease-drug indications spanning 68 diseases and evaluate in silico our approach against the most widely used transcriptomics-based matching baselines, CMap and the Characteristic Direction. Our results show a more than 200\% improvement over both baselines in terms of standard retrieval metrics. We further showcase our model's ability to capture different genes' expressions interactions among drugs and diseases. We provide our trained models, data and code to predict with them at \url{https://github.com/healx/dgem-nn-public}}.


\maketitle

\section{Introduction}\label{sec1}

The standard paradigm of drug discovery for a specific disease involves determining one or multiple target proteins that are disease-modifying and then screening or designing small molecules that bind, i.e. modulate, the target(s). This process is hindered by multiple factors, such as discovering druggable targets or avoiding off target drug effects, and is in part responsible for the notoriously long development time and low success rate in getting drugs to the clinic.

An alternative to the above, hypothesis-driven paradigm, is conversely a hypothesis-free strategy where we seek to find drugs for a disease based on phenotypic screens, usually omics or cell imaging data. This approach is particularly suitable for rare diseases or diseases with complex etiology, where limited knowledge of the disease biology might make target discovery unfeasible. 

The focus of this work is specifically a transcriptomics-driven approach for matching drugs to diseases. The Connectivity map initiative \cite{lamb2006connectivity} and its subsequent L1000 version \cite{subramanian2017next} have generated a large dataset of drug and gene perturbations, that paved the way to connect transcriptomically diseases, genes and drugs, and has been successfully applied to identify new active molecules \cite{hieronymus2006gene, wei2006gene, de2018high} drug repurposing \cite{aliper2016deep, iorio2013transcriptional} and mechanism of action \cite{iwata2017elucidating, wacker2012using} among others. In order to match a query disease transcriptomic profile against the above datasets, the authors relied on a Kolmogorov-Smirnov similarity score. Several improvements on the initial matching algorithm have been proposed in the literature \cite{zhang2008simple, pacini2013dvd, cheng2014systematic, duan2016l1000cds2}, with all of them employing statistical methods or analyses to match drugs to disease profiles. Of note, all methods attempt to match drugs to diseases based on per-gene effects, e.g. matching the effect of geneA in the disease profile vs the effect of geneA in the drug profile. Furthermore, genes are included or omitted in the calculations based on statistical properties, such as statistical significance of the changes. 

Unfortunately, there are a number of issues inherent to the nature of the data that make it challenging getting accurate predictions with simple statistical methods. First, transcriptomics data have by default many missing values which makes a per gene comparison of disease and drug profiles challenging. This problem could perhaps be addressed by modelling the complex correlations of how genes are expressed, e.g., how an upregulating effect of geneX in the disease is linked to a downregulating effect from the drug on geneY. Additionally, these methods treat all gene transcripts as equal, although some genes play a more significant biological role than others and therefore a small deregulation of them might have more significant effects. Another important factor relates to the many sources of noise in transcriptomics measurements that can be only partially addressed with traditional statistical methods. Lastly, and linked to the previous point, we shouldn't disregard that most often, when trying to match a drug to a disease in this setting, we employ the transcriptomic profile of a drug measured on a cell line and the respective profile of a disease measured on totally different cells, either human or animal tissue samples.

Here, we aim to address the above limitations by developing a deep learning neural network architecture that can learn, given one or multiple disease contrasts, to predict whether a disease is likely to be treated by a drug, by comparing their gene expression profiles. Our proposed approach learns to correlate disease vs drug transcriptomic changes even if the disease and the drug do not share the same significantly dysregulated genes, as we show in Section \ref{subsection: use case}. Importantly, compared against two widely employed baselines our approach achieves a 200\% improvement over them in terms of hits@k, i.e., ranking known treatments in the top positions, on a dataset of 68 diseases and 673 approved indications.

Specifically, the contributions of this work can be summarized as follows:
\begin{itemize}
    \item We assemble a training and an evaluation dataset by combining disease and drugs transcriptomic profiles and labelling them with their treatment approval.
    \item We develop a neural network architecture that is trained on the above dataset and can, at inference, predict for a given pair of drug and disease contrasts, the likelihood of the drug being a treatment for the disease.
    \item We evaluate empirically our model against two popular baselines for drug-disease transcriptomics-based matching, showing very large improvements over both of them.
    \item Lastly, we further analyze our model's results showing that it can learn how transcriptomic changes across different genes in the disease and the drug can interact and employ this knowledge to predict correctly treatments for diseases.

\end{itemize}

A number of past works developed machine learning models leveraging the L1000 data, such as \citet{pham2021deep} that trained a deep learning architecture to predict trascriptomic profiles of new compounds or \citet{mendez2020novo} who conditioned Generative Adversarial Networks on gene knockouts to design new molecules that would achieve the same effect. Nevertheless, and to the best of our knowledge, our work is the first to present a model that learns directly and jointly from disease and drug transcriptomic profiles in order to come up with new drug indications for diseases.

\section{Methods}\label{sec:methods}

In this section we start by defining the task and subsequently present the datasets, the model architecture and the baselines.

\subsection{Task Definition}

Let $D$ be the set of diseases and $C$ the set of compounds in our dataset. An approved treatment indication is a pair of $(c_i, d_j)$ with $c_i \in C$ and $d_j \in D$ respectively, $c_i$ being an approved treatment for $d_j$. 

Furthermore, we assume that each disease $d_j$ is described by one or multiple two dimensional vectors $v_{dj}$ of size $\|T\|$ where $T$ is the set of transcripts. These vectors represent transcriptomic profile(s) of $d_j$. Each of these profiles have been produced with differential gene expression tools such as Sleuth \cite{pimentel2017differential} or DESeq2 \cite{love2014moderated}, comparing a diseased human or animal model tissue sample against a healthy matching sample. Each position in these vectors contains the log fold change of the respective transcript along with a p-value which signals if the change is statistically significant or not. Of note, since the process of measuring differential gene expression is inherently noisy and influenced by various factors (such as sex or age of the donor), we ideally want each disease to be described by multiple samples, i.e., transcriptomic profiles. In the following, we will refer to each of these profiles coming from different samples, as disease contrasts.

Similarly, each compound $c_i$ is described by one or multiple vectors $v_{ci}$ of size $\|L\|$, with each element of the vector representing the log fold change of the respective transcript. Each of these vectors represents the measured transcriptomic profile when treating a specific cell line with a specific drug concentration. It is important again to note that the measurement process is noisy and furthermore influenced by the specific cell line, therefore we ideally want multiple measurements across different cell lines to allow our neural network to learn useful patterns. In the following, we will call each of these drug transcriptomic profiles as a drug contrasts.

With the above definitions, we cast the learning task to a binary classification problem. Specifically, we design a neural network architecture which takes as input pairs of $(v_{ci}, v_{dj})$ vectors and learns to classify them as either 1 or 0 based on the likelihood that $(c_i, d_j)$ is an indication. We discuss in Section \ref{section:dgem-nn} the design choices for the aforementioned architecture and the rationale behind it.

\subsection{Dataset}

In order to train and evaluate our model, we construct a dataset by employing three sources: a) a list of known indications for diseases b) a list of disease contrasts and c) a list of drug contrasts.

\begin{table}[h]
\begin{center}
\caption{Statistics of the training and evaluation datasets.}\label{tbl: datasets statistics}%
\resizebox{\columnwidth}{!}{%
\begin{tabular}{cccccc}
\toprule
Dataset & Diseases & \begin{tabular}{@{}c@{}}Diseases\\Contrasts\end{tabular}&Drugs&\begin{tabular}{@{}c@{}}Drugs \\Contrasts\end{tabular}& Total positives\\
\midrule
Training\&Validation& 207   & 1,285  & 5,554 & 33,069 & 244,339\\
Test& 68 & 975 & 673 & 4,894 & 186,050\\
\botrule
\end{tabular}
}
\end{center}
\end{table}

For the disease contrasts we employ proprietary data encompassing 2,260 contrasts from 275 diseases. As we mentioned, each of these contrasts has a vector size $\|T\|$. We set $\|T\| = 30374$ for our data. For each position we have both the log fold change and the p-value available (see previous Section). We process the data by setting to zero, for each gene-position in each vector, any log fold value that has a p-value larger than 0.3. We then drop the p-values and obtain this way a mostly sparse, one dimensional vector of size $\|T\|$ for each disease contrast.

For the above diseases, we collect known indications from our internal Knowledge Base, resulting in 3,725 indications. For the drug contrasts we employ the L1000 dataset which encompasses a list of 39k drug perturbations across different cell lines, dosages and time measurements totalling more than 1.2M perturbations. We consider the 10 most frequent cell lines and dosages in the order of $10^1 \mu M$, and only measurements at 24h. This filtering results in a dataset of 5,554 drugs and more than 33k respective contrasts in total, across the different cell lines. Each of these contrasts has a size $\|L\|=12,328$. We split the resulting dataset into a training and test set. The statistics of the datasets are summarized in Table \ref{tbl: datasets statistics}. To simulate a drug discovery scenario where there is no known treatment for the target disease, our split makes sure that the test set doesn't contain any of the diseases included in the training set, but we might have drugs appearing in both sets.

\subsection{Differential Gene Expression Matching Neural Network (DGEM-NN)}\label{section:dgem-nn}

\begin{figure}[h]
\centering
\includegraphics[width=0.9\textwidth]{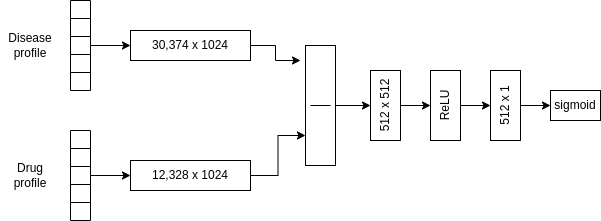}
\caption{DGEM-NN architecture}\label{figure: architecture}
\end{figure}

As already noted, our neural network architecture takes as inputs the $(v_{ci}, v_{dj})$ vectors and learns a binary classification task. Specifically, each of the input is passed through a disease encoder and a drug encoder respectively which are linear layers with ReLU activation, outputting dense vectors of 1,024 elements. These vectors are concatenated and passed through another two linear layers with a ReLU activation in between and a sigmoid activation on top. Figure \ref{figure: architecture} illustrates the architecture and inputs of the network. 

We should clarify that the dataset that we constructed above contains for each drug-disease pair, all available drug-disease trascriptomic profiles combinations. For instance, if a disease has five different contrasts and we have six different contrasts for a drug across different cell lines, then we include in total thirty training samples for that pair. Our aim in doing so is that the model will learn a representation for the disease that is invariant to the noise and other sample-specific factors (sex, age, etc). Similarly, we want to learn a drug representation that is independent of noise and cell line.

As mentioned, we cast this problem to a binary classification task and therefore need to also provide the model with negative examples. Since there are no explicit negatives available, we resort to a negative sampling approach, commonly followed in similar positive-unlabelled tasks: for each positive pair, we sample five "negatives", in reality unknown, pairs of drugs and diseases. In order to minimize the possibility of sampling a true positive but unknown indication, we make sure that for each negative sample disease-drug pair, the drug is both structurally and transcriptomically different to any known indication for the disease. We test structural similarity in terms of the compound molecular fingerprints with the FPSim2 package\footnote{\url{https://github.com/chembl/FPSim2}} and transcriptomic similarity in terms of cosine similarity of the contrasts of the drugs.

We consider beneficial here to share also a number of failed or suboptimal architectures and model choices that we experimented with in the early phase of this work. One of our initial choices was to include in our disease inputs the p-values of the statistical significance tests across transcripts, which are commonly provided in trascriptomic profiles. In that setting, the disease encoder was a convolutional neural network. After observing a suboptimal performance in intial experiments, we chose to omit these inputs and, as mentioned, set to zero any gene transcript changes that had a p-value larger than 0.3. A few other design choices included adding molecular information for the compounds in the form of Morgan fingerprints and modelling cell lines with a separate cell line encoder, both of which didn't yield favourable results.

\subsection{Baselines}

We compare DGEM-NN against two established and widely used baselines, the CMap algorithm presented by \citet{lamb2006connectivity} and \citet{subramanian2017next} and the Characteristic Direction method presented by \citet{duan2016l1000cds2}. 

The first method, also called weighted connectivity score (WTCS), represents a non-parametric, similarity measure based on the weighted Kolmogorov-Smirnov enrichment statistic (ES) \cite{subramanian2005gene}. WTCS is a composite, bidirectional version of ES. For a given query gene set pair ($q_{up}$, $q_{down}$ ) and a reference signature $r$, WTCS is computed as follows:
\begin{equation}
    w_{q,r} = \begin{cases}
(ES_{up}-ES_{down})/2 & \text{if $sgn(ES_{up}) \neq sgn(ES_{down})$}\\
0 & \text{otherwise} 
\end{cases}
\end{equation}
where $ES_{up}$ is the enrichment of $q_{up}$ in $r$ and $ES_{down}$ is the enrichment of $q_{down}$ in $r$. WTCS ranges between -1 and 1. Commonly, WCTS is averaged across cell lines and perturbagen types and a normalized connectivity score is computed. For more details, we refer the interested reader to the project's webpage\footnote{\url{https://clue.io/connectopedia/pdf/cmap_algorithms}}. In our experiments, we employ a proprietary implementation of the CMap algorithm.

The second baseline, Characteristic Direction (CD), was first introduced by \citet{clark2014characteristic}. It represents a multivariate method to compute signatures, giving less weight to individual genes that display a large change in magnitude when comparing two conditions, for example, comparing gene expression from drug-treated cells with control cells. Instead, the CD method gives more weight to genes that move together in the same direction across repeats and therefore a gene that changes less but ‘moves‘ together with a large group of other genes may be scored higher than a gene that changed more in overall magnitude. To compare against DGEM-NN, we queried the relevant L1000CDS2 API\footnote{\url{https://maayanlab.cloud/L1000CDS2/help}} for each of the contrasts in our test set and subsequently followed the same post-processing procedure as with DGEM-NN. We provide details regarding postprocessing in the following Section.

\section{Results}

In this section we first compare DGEM-NN against the aforementioned baselines and then present an experiment highlighting DGEM-NN's capability in understanding how different gene effects are related across diseases and drugs.

\subsection{Comparing DGEM-NN against the baselines}

We split our training set in a 80\%-20\% training-validation split, making sure our validation set diseases don't overlap with the ones from the training set. We then train DGEM-NN employing the Adam optimizer with a learning rate of $0.001$, a batch size of $256$ and using the binary cross-entropy as our loss function, for up to 10 epochs, keeping the best model on the validation set in terms of the Area under the ROC curve metric. Of note, we are using these parameter values after doing a parameters grid search on the validation set. During inference, we query our model with each of the test set's transcriptomic profiles and obtain a ranking of drug perturbations for each of them, with the model's probability associated with each.

For the baselines, we either run our CMap implementation or query the L1000CDS2 API with each of the disease contrasts of our test set.

All three methods return a ranking of drug contrasts for each disease contrast. In order to obtain a ranking of drugs for each disease, we follow the same postprocessing across all methods. Specifically, we keep at most the top 500 predicted drug contrasts for each disease contrast and average over them to obtain an aggregate score for each drug-disease pair by which we rank predicted drugs for each disease.

\begin{table}[h]
\begin{center}
\caption{Methods evaluation on our test set. Results in bold represent statistically significant results against the baselines.}\label{tbl: comparison}%
\begin{tabular}{cccc}
\toprule
Method & hits@50 & hits@100 & MRR \\
\midrule
CMap \cite{lamb2006connectivity} &0.16 &0.26 &0.02 \\
Characteristic Direction (CD) \cite{duan2016l1000cds2}&0.21 &0.38 &0.02 \\
DGEM-NN (1 model) &\textbf{0.36} &\textbf{0.51} & \textbf{0.04}\\
\botrule
DGEM-NN ensemble(voting, 20 models) & \textbf{0.39}&\textbf{0.55}&\textbf{0.10} \\
\botrule
\end{tabular}
\end{center}
\end{table}

In order to evaluate the predictions we compare them against the known indications in the test set. We need to note that this evaluation has limitations since we essentially evaluate how many known treatments each method has ranked on the top positions, but we don't necessarily know all possible treatments for a disease. This is a standard situation in ranking tasks such as search engines, recommendation systems or knowledge graph completion methods, therefore we compare all methods in terms of standard ranking metrics, hits@k\footnote{\url{https://pykeen.readthedocs.io/en/stable/api/pykeen.metrics.ranking.HitsAtK.html}} and Mean Reciprocal Rank (MRR)\footnote{\url{https://pykeen.readthedocs.io/en/stable/api/pykeen.metrics.ranking.InverseHarmonicMeanRank.html}}\cite{hoyt2022unified}. We evaluate at hits@50 and hits@100 in order to mitigate for diseases that have a large number of treatments and for the aforementioned observation, that the algorithms might propose treatments which are valid but unknown and thus considered false negatives. In Table \ref{tbl: comparison} we summarize the results when comparing the baselines against a DGEM-NN model (average results across five runs) and an ensemble of 20 DGEM-NN models. We observe a very large improvement over both of the baselines for the DGEM-NN method, both for the single model and the ensemble.

\begin{table}[h]
\begin{center}
\caption{Evaluation on our test set, per disease area in terms of hits@100. Results in bold represent statistically significant results against the baselines.}\label{tbl:disease-areas}%
\begin{tabular}{ccccc}
\toprule
Disease Area &\# of Diseases & CMap & CD & DGEM-NN ensemble \\
\midrule
Autoimmune &7&0.31&0.44&\textbf{0.92} \\
Cancer &5&0.55&0.57&\textbf{0.72} \\
Cardiovascular &5&0.30&\textbf{1.0}&0.26 \\
Infection Bacterial &7&0.09&0.4&\textbf{0.72}\\
Infection Viral &2&0.13&0.22&\textbf{0.33}\\
Inflammation &10&0.31&0.17&\textbf{0.59}\\
Kidney &2&0.17&1.0&\textbf{0.23}\\
Neurodegenerative &3&0.12&0.33&\textbf{0.89} \\
Nutritional/metabolic &6&\textbf{0.39}&0.19&0.25 \\
Skin &6&0.1&0.28& \textbf{0.64}\\
Psychiatric &6&\textbf{0.42}&0.1&0.33\\
Respiratory &3&0.13&0.21&\textbf{0.35}\\
Other &6&0.17&0.4&\textbf{0.45}\\
\botrule
\end{tabular}
\end{center}
\end{table}

To further provide insights into these results and since our test dataset spans across 13 disease areas, we further analyze results across them. In Table \ref{tbl:disease-areas} we report the results in terms of hits@100. Interestingly, we observe large fluctuations in terms of hits@k across the disease areas, which can be attributed to an extent to the technical noise in measurements across the different disease but most importantly likely to the fact that not all disease areas might benefit equally from transcriptomics based drug matching. In terms of the different methods, we observe again a steady and significant advantage of our method against the baselines across the majority of diseases.

\subsection{DGEM-NN Captures Different Genes Interactions between Drugs and Diseases}\label{subsection: use case}

\begin{table}[h]
\begin{center}
\caption{Examples where DGEM-NN predicts correctly a drug for a disease, even when there are no common genes with significant transcriptomic changes measured, by implicitly learning interactions across genes. The "drug/disease genes" columns present significantly dysregulated genes in the drug and disease contrasts that are involved in the common pathway.}\label{tbl:use case}%
\resizebox{\columnwidth}{!}{
\begin{tabular}{ccccc}
\toprule
Disease&Drug& Common Pathways& Drug genes & Disease genes\\
\midrule
\begin{tabular}{@{}c@{}}Systemic Lupus\\Erythymatosus\end{tabular}&Dexamethasone&\begin{tabular}{@{}c@{}}collagen fibril \\organization (GO:0030199)\end{tabular}&\begin{tabular}{@{}c@{}}COL4A2;COL4A6;\\COL4A5;PLOD2;\\ITGA6;PPIB\end{tabular}&COL21A1\\
Rheumatoid Arthritis&Cyclosporin A&\begin{tabular}{@{}c@{}}actin-myosin filament\\sliding (GO:0033275)\end{tabular}&\begin{tabular}{@{}c@{}}MYBPC3;TPM4;\\MYL2;TTN\end{tabular}&\begin{tabular}{@{}c@{}}ACTN3;ACTN2;\\TNNI1;MYH6\end{tabular}\\
Alzheimer's Disease&Memantine&\begin{tabular}{@{}c@{}}regulation of protein\\polymerization (GO:0032271)\end{tabular}&\begin{tabular}{@{}c@{}}ABITRAM;RASA1;\\ABL1\end{tabular}&PFN2\\
Psoriasis&Prednisone&\begin{tabular}{@{}c@{}}protein deubiquitination\\(GO:0016579)\end{tabular}&\begin{tabular}{@{}c@{}}PSMA5;OTUB2;\\ATXN3;BECN1;\\PSMC5;USP15;\\PSMC2;PTEN
\end{tabular}&\begin{tabular}{@{}c@{}}USP17L15;USP17L5;\\USP17L17;USP17L19;\\USP17L20;USP17L12;\\USP17L11;USP17L24\end{tabular}\\
\botrule
\end{tabular}
}
\end{center}
\end{table}

Since DGEM-NN shows in silico a considerable improvement against the baselines, at least in terms of ranking known treatments in the top positions, we sought to further analyze which factors drive this increase in coverage. As mentioned, in the context of transcriptomics data many values are missing or noisy across genes. Our data are reminiscent of these issues, since we have sparse representations of different dimensionalities for the drugs and diseases, 12,328 and 30,374 respectively. One hypothesis that we would like to test is whether DGEM-NN can correlate differences in the expression of dysregulated genes in a disease contrast against the ones in a drug contrast at a pathway level, rather than at an individual gene level, and hence being able to predict a correct treatment, even though a disease-drug contrast pair might not have any common dysregulated genes. To that end we designed the following experiment:

We picked randomly ten diseases and considered contrasts of these diseases for which the model predicted correctly known drugs as treatments, but which didn't share any common genes significantly dysregulated. In these cases, we selected the top 250 features-genes for each of the drug and disease inputs and employed Enrichr \cite{chen2013enrichr} through the GSEApy package \cite{fang2023gseapy} and using the Gene Ontology Biological Process library 2021 edition, to find if there were any common biological pathways between the drug and the disease dysregulated genes. Out of 3,392 total correct predictions for the 342 diseases' contrasts there are 175 cases where the two pairs don't share any common gene significantly dysregulated and in 144 of them they do share a common pathway. In Table \ref{tbl:use case} we illustrate a few representative examples (we include all 144 cases in the public repository). Interestingly, the common pathways are known to be affected in the respective diseases in the literature, for instance Lupus is a collagen disease, protein polymerization is affected in Alzheimer's \cite{m2014protein} and protein deubiquitination is affected in Psoriasis \cite{yang2018ubiquitination}.

These results suggest that our model might be capable to an extent, without any explicit knowledge injected to it, to understand how transcriptomic changes across different genes in the disease and the drug can interact at a pathway rather than gene level and employ that knowledge to predict correctly treatments for diseases. It should be noted that this type of information provides a valuable insight to a highly sought aspect of virtual screening, which is revealing the mechanism of action by which drugs act on diseased cells.

Since these learned correlations offer valuable insights to the model's decisions and hence an interpretability aspect, in our publicly shared code we include this feature when conducting predictions.
\section{Conclusions}\label{conclusions}

In this work we presented a deep learning approach to finding drugs for diseases based on the relevant transcriptomics profiles. Our method, DGEM-NN, overcomes several limitations of previously developed statistical methods to perform transcriptomics-based drug matching and achieves a 200\% improvement over the two most widely used baselines in terms of hits@k, i.e., ranking known treatments in the top positions, on a dataset of 68 diseases and 673 approved indications. We furthermore show that our model is, at least to an extent, able to recommend a drug for a disease based on shared biological pathways, even if the two transcriptomic profiles don't share any genes with significant changes.

Since our approach is mainly agnostic to the perturbation type, it can readily be extended in applications such as target identification, where we would match the transcriptomic profile coming from a gene knockout to a disease. Also, our approach can be coupled with recent approaches trying to predict transcriptomic profiles for novel compounds \cite{pham2021deep} in order to build an end to end virtual screening model that gets as input disease transcriptomic data and outputs novel compounds that potentially reverse the disease.

\section{Models, Code and Data availability}

The full training and test data is proprietary and therefore not released. Nevertheless, we provide publicly our trained models along with the necessary code and processed L1000 data to infer with it, to allow further evaluation of our approach.

\backmatter

\bmhead{Acknowledgments}

We would like to thank our colleagues, Azedine Zoufir, Daniel Mason and Ian Roberts who gave valuable feedback and suggestions throughout this work.

\bibliography{main}
\bibliographystyle{sn-basic}
\end{document}